# Fiber-integrated silicon carbide silicon vacancy-based magnetometer

Wei-Ke Quan[1,†], Lin Liu[2,†], Qin-Yue Luo[1], Xiao-Di Liu[2,*], Jun-Feng Wang[1,*]

[1]*College of Physics, Sichuan University, Chengdu, Sichuan 610064, People's Republic of China*

[2]*Key Laboratory of Materials Physics, Institute of Solid State Physics, HFIPS, Chinese Academy of Sciences, Hefei 230031, China*

[†]These authors contributed equally to the work.

[*]Corresponding author: xiaodi@issp.ac.cn, jfwang@scu.edu.cn

**Abstract**

**Silicon vacancy in silicon carbide has drawn much attention for various quantum sensing. However, most of the previous experiments are achieved using confocal scanning systems, which limit its applications in practical applications. In this work, we demonstrate a compact fiber-integrated silicon carbide silicon vacancy-based vector magnetometer at room temperature. First, we effectively couple the silicon vacancy in a tiny silicon carbide slice to an optical fiber tip and realize the readout of the spin signal through the fiber at the same time. We then study the optically detected magnetic resonance spectra at different laser and microwave powers, obtaining an optimized magnetic field sensitivity of 12.3 μT/Hz$^{1/2}$. Based on this, the magnetometer is performed to measure the strength and polar angle of an external magnetic field, respectively. Through these experiments, we have paved the way for fiber-integrated silicon vacancy-based magnetometer applications in practical environments such as geophysics and biomedical sensing.**

In recent years, color centers in silicon carbide (SiC) have been considered promising platforms in quantum technologies due to their excellent spin and optical properties[1-19]. Various high bright single-photon emitters and diodes have been identified and fabricated in SiC[5-9]. Moreover, there are also several solid-state spin qubits including silicon vacancy[3], divacancy[1,2,4] and nitrogen-vacancy centers[9-11] in different SiC polytypes. The spin state of the spin qubits can be initialized, read out by optics, and controlled by microwave even at room temperature[3,4]. Silicon vacancy is a point defect containing a silicon vacancy in SiC crystal lattice. As a spin qubit, it has some prominent properties such as its near-infrared wavelength and long coherence time at room temperature[3,12]. Due to these advantages, it has been applied in quantum photonics[8,13], high-fidelity spin-optical interface[14], quantum information process[15], and quantum sensing such as magnetic field[16,17], mechanics[18], temperature[19] and so on. In the case of quantum sensing applications, nanotesla magnetometry has been realized using high concentration silicon vacancy in 4H-SiC[16,17]. However, the magnetometer is realized based on the macroscopic confocal system, which is not suitable for practical application[16,17].

To release the capability of the magnetometer and make them compatible with the requirements of practical environments, the silicon vacancy magnetometer needs to be integrated with optical fiber[20-26]. Compared with NV centers in diamond, silicon vacancy in SiC has several advantages[1,3,8,12-14]. First, SiC is a mature semiconductor material with important technological advantages such as inch-scale growth and micro-nano fabrication technologies. SiC-based quantum technologies are easily compatible with the modern semiconductor industry and can be widely used[1-3]. Second, the wavelength of the silicon vacancy is in near-infrared, making it more suitable for communication in fiber networks and longer distance distributed fiber magnetometry[1,3,8]. Finally, it has only one direction (perpendicular to the SiC surface) and the corresponding optically detected magnetic resonance (ODMR) spectra has only two peaks which is beneficial for calibrating and sensing the magnitude and direction of the magnetic field[1,3,8]. Moreover, its zero-field-splitting (ZFS) is temperature independent[12,19], which is essential for magnetic sensing in practical environments.

In this work, we realize a fiber-integrated silicon carbide silicon vacancy-based vector magnetometer at room temperature. First, we paste a SiC sample with a diameter of 100 μm on a fiber tip and realize an efficient coupling of the silicon vacancy and the fiber. At the same time, a few rounds of copper wires as a microwave antenna are moved close to the SiC sample to control the spin state of the silicon vacancy. We then investigate the ODMR as a function of laser and microwave powers, obtaining an optimized magnetic sensing sensitivity of 12.3 μT/Hz$^{1/2}$. To demonstrate that it could be used as a vector magnetometer, we finally measure the ODMR as the magnetic strength and polar angle of a magnetic field vary, respectively. These results pave the way to fiber-integrated silicon vacancy-based magnetometer being used in practical magnetic field sensing.

In these experiments, we use a high concentration ensemble of silicon vacancies (V2 center) in a 4H-SiC sample. The sample is implanted by 20 keV He ions with does of $1\times10^{13}$/cm$^2$, then annealed at 500 °C for 2 h[27]. The depth of the silicon vacancy is about 120 nm[27]. To paste on the optical fiber end face, we then mechanically polish the sample to a thickness of 50 μm for pasting on the optical fiber end face. As depicted in Fig.1a, we build a hybrid fiber-optical system and microwave (MW) system to measure the silicon vacancies signal. A 785 nm laser is collimated by the lens to couple with a multimode fiber connected to the compact cage optical system. The laser is reflected by a dichroic mirror and mirror, and coupled into a multimode fiber (50 μm core diameter, 0.22 NA). The laser is focused on the sample by a multimode fiber, meantime, the fluorescence of the silicon vacancies is collected by the same multimode fiber. The fluorescence signal is filtered is by an 850-nm long-pass filter and collected to a photoreceiver (Femto, OE-200-Si)[12,27]. For the microwave system, the MW generator and MW switch are controlled by a PC and amplified by a high-power MW amplifier (Mini-circuits, ZHL-25W-272+). Microwaves signal are delivered to the lab-winded copper coils to excite the SiC sample. A magnet is used to provide a proper external magnetic field in the experiments. The standard lock-in methods are used to measure the ODMR signal[1,12,27]. As shown in Fig. 1b and 1c, the cladding and the core of the multimode fiber are 2500 μm and 120 μm in diameter, respectively. A SiC sample with the size of about 120×120×50 μm³ is pasted on the fiber core using UV curing glue. As presented in Fig. 1d and 1e, a few rounds of copper coils are placed close to the SiC sample to efficiently transmit MW signals to control the spin of silicon vacancies.

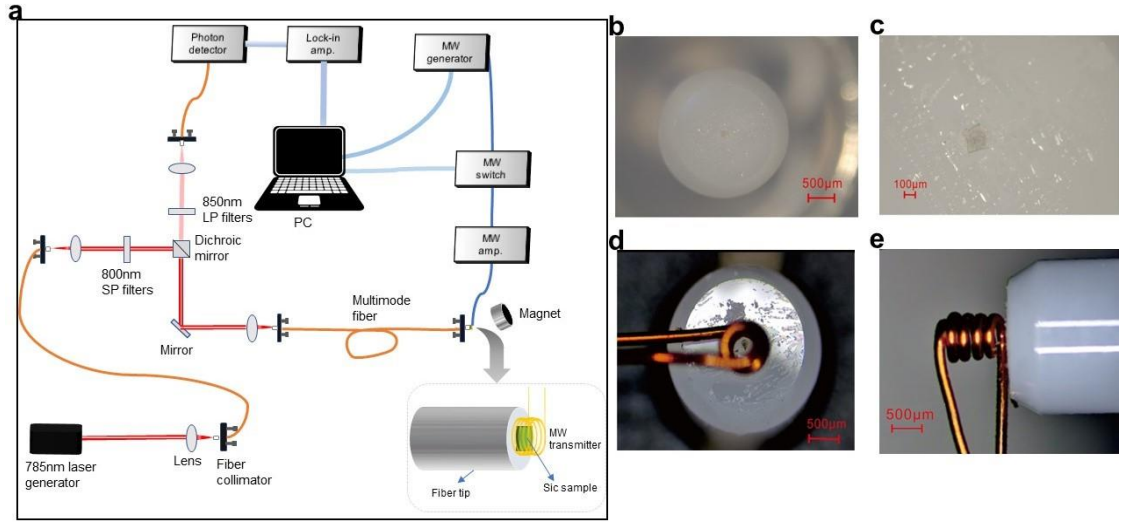

**Figure 1** Experimental set-up and SiC sample. (a) Schematic diagram of the experimental setup. The orange curve represents multimode fiber. The exciting laser is represented by a darker red line while PL signals are lighter red as of which wavelength is more infrared. Light blue lines represent electrical signal transmission, and dark blue lines represent microwave transmission. (b) and (c) Optical images of the fiber tip adhered with a SiC sample about 100 μm in diameter. (d) and (e) Front and side optical images of the microwave coil and sample, respectively.

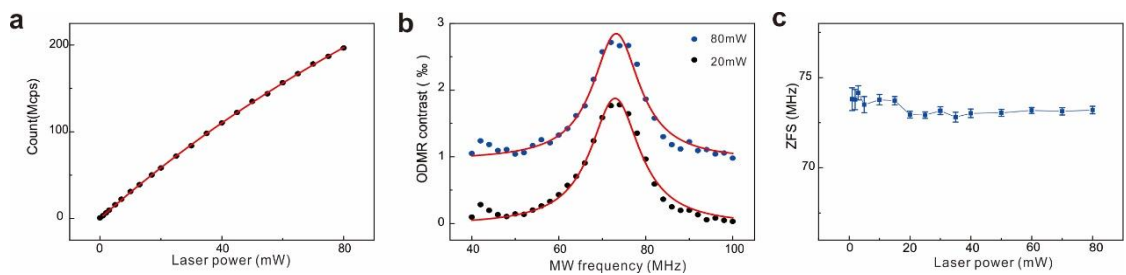

**Figure 2** Saturation curve of silicon vacancy. (a) Saturation curve of silicon vacancy counts versus laser power, the red line is fitting to the data. (b) ODMR at different laser powers, the red curves are the Lorentz fitting of data. (c) ZFS values as a function of the laser power. The error bars are due to standard deviations of fitting to data.

Without using a complicated confocal system, the fiber-integrated magnetometer shows good environmental adaptability. The spatial resolution depends on the size of the SiC sample. For this magnetometer, SiC samples with a diameter of about 100 μm. To characterize the photoexcited properties of the sample, PL counts at different laser powers are presented in Fig. 2a. The data are fit using the function $I(P) = I_s/(1+P_0/P)$, where $I_s$ is the maximum count, and $P_0$ is the saturation power. Inferred from the fitting, the saturation count $I_s$ is 935 Mcps and the saturation power $P_0$ is 300 mW, respectively. The high saturation count demonstrates the efficient coupling of the silicon vacancy and the multimode fiber. In the experiment, we use the ODMR methods to measure the magnetic field[16,17,28]. The electronic spin Hamiltonian of silicon vacancy in 4H-SiC is given as[3]:

$$H=D[S_z^2-S(S+1)/3] + g\mu_B B_0[S_z \cos(\theta)+S_x \sin(\theta)] \quad (1)$$

where S is 3/2, g is the Landé g-factor, $\mu_B$ is the Bohr magneton, $B_0$ and $\theta$ are the strength and polar angle of the applied static magnetic field, respectively. D represents the zero-field-splitting (ZFS) parament which is about 35 MHz of silicon vacancy in 4H-SiC. Without an external magnetic field, the two energy levels $|\pm 3/2\rangle$ are degenerate as well as the $|\pm 1/2\rangle$, inducing a resonance frequency of 2D. Under the external magnetic field, the spin degenerates have Zeeman splitting, resulting in two dipole-allowed transitions, $v_1$ ($|-1/2\rangle \leftrightarrow |-3/2\rangle$) and $v_2$ ($|1/2\rangle \leftrightarrow |3/2\rangle$)[3].

For NV center in diamond, with high laser power, the laser will heat the diamond samples on the fiber tips[20,21,26]. Since the ZFS values of the NV centers decrease as the temperature increase, the heat effect can result in the fluctuations and shifts of ODMR resonance frequency. To keep a constant temperature, the exciting laser power has to be set bellow 10 mW, which will limit its sensitivity and applications in practical environments[20,26]. For silicon vacancy in 4H-SiC, the D is temperature-independent, so the ODMR peaks are not affected by laser thermal effects, which is crucial for practical applications and high sensitivity[12]. To confirm this vital property, we performe ODMR experiments on the same sample at different laser powers. OMDR measurements at two different laser powers are shown in Fig. 2b. By Lorentzian function fitting, the ZFS value at 20 mW (73.2 MHz) is consistent with that at 80 mW (73.0 MHz). Figure 2c shows how the ZFS values vary with laser powers. Unlike NV centers in diamonds, ZFS values of silicon vacancy are generally unchanged with the laser powers ranging from 1 mW to 80 mW[26]. The temperature stability is vital for ODMR-based magnetometers as the thermal effects of the laser are difficult to eliminate and control, which makes it a prominent advantage of the silicon vacancy-based magnetometer.

Since both the laser and MW powers can broaden the linewidth of ODMR peaks[12,28-30], which will affect the magnetic field sensing sensitivity. To obtain optimal measurement settings for the highest magnetic field sensing sensitivity[17,29], it is necessary to explore the effect of experimental condition settings on ODMR parameters. Figures 3a and 3b present the ODMR contrast, ODMR full width at half maximum (FWHM) as a function of the laser power at fixed MW power of 18 dBm, respectively. Unlike the NV centers in diamond, the silicon vacancy ODMR contrast and FWHM are fixed at about 1.8‰ and 13 MHz as the laser power varies respectively, which is consistent with previous results[30]. As a key metric of magnetometer performance, ODMR-based magnetic field sensitivity can be approximately represented as $\eta_B \approx 0.77 \frac{h}{g\mu_B} \frac{\Delta v}{C\sqrt{R}}$, where R is the rate of detected photons, C is the ODMR contrast, h stands for Planck's constant, and $\Delta v$ is the ODMR width[28]. Figure 3c shows the magnetic field sensitivity as a function of the laser power, which tells that the sensitivity increases from 118.5 μT/Hz$^{1/2}$ to 13.5 μT/Hz$^{1/2}$ as the laser increase from 1 mW to 85 mW. Meanwhile, Figures 3d and 3e show how contrast and FWHM vary with MW Power with laser power fixed at 60 mW. It illustrates that both the ODMR contrast and FWHM markedly increase with MW power in the increasing from 3 to 23 dBm, which is in accord with previous color centers in SiC[17,29,30], and diamond[28]. The sensitivity as a function of MW powers is shown in Fig. 3f. The optimal sensitivity can be obtained at a microwave power of 19 dBm. The highest measured sensitivity is about 13.4 μT/Hz$^{1/2}$ at around 19 dBm MW power. It can be further increased to 12.3 and 5.4 μT/Hz$^{1/2}$ at the highest experimental laser power (85 mW) and saturation count, respectively.

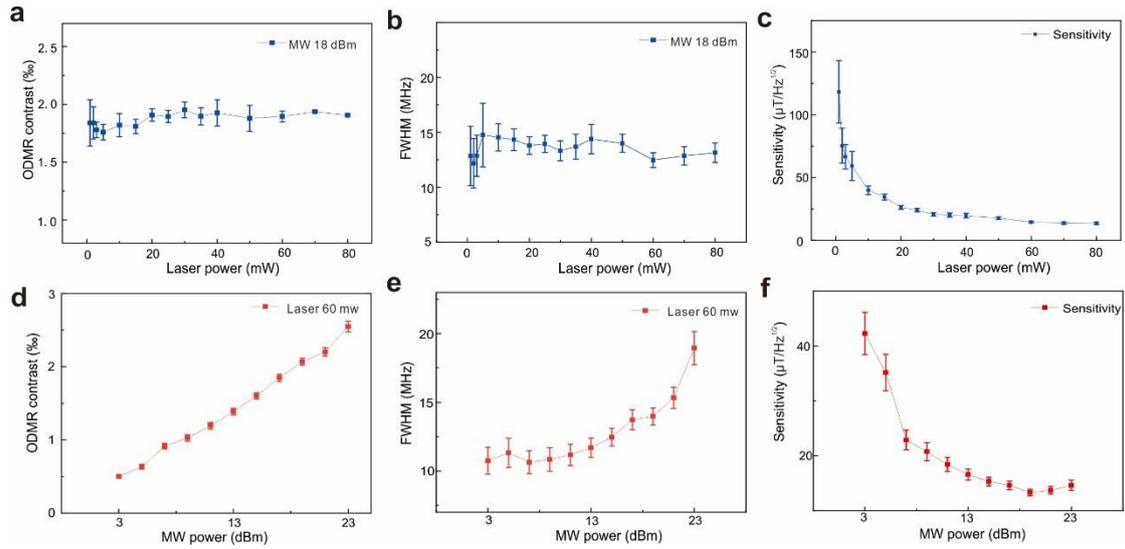

**Figure 3** Optimization of the magnetic field sensing sensitivity. (a) and (b) ODMR contrast and line width as a function of the laser power with the microwave power of 18 dbm, respectively. (c) Magnetic field sensitivity with respect to laser power. (d) ODMR contrast and line width as a function of the MW power at 60 mW laser power, respectively. (f) Magnetic field sensitivity with respect to MW power. Error bars are the standard deviations of the corresponding fittings.

After the laser and MW powers are optimized, we then measure the strength and polar angle of an applied external magnetic field to confirm the applicability and accuracy of the fiber-integrated vector magnetometer with laser power of 60 mW and MW power of 18 dBm. First, we measure ODMR spectra under the c-axis magnetic field ($\theta=0$). Figure 4a presents three representative ODMR spectra at different magnetic fields. The resonance frequencies increase under larger magnetic fields. The summary of the resonance frequencies ($\nu_1$ and $\nu_2$) under different magnitudes of the magnetic field is represented in Fig. 4b, which is consistent with theoretical results by Eq. 1 (Red lines) in the range of magnetic field increase from 0 Gs to 120 Gs. Furthermore, to check the response of the vector magnetometer to the magnetic field angle $\theta$, we fix the $B_0$ at 60 Gs and change the polar angle $\theta$ from 0 º to 90 º by rotating the magnet. Figure 4c presents the resonance frequencies $\nu_1$ and $\nu_2$ as a function of polar angle $\theta$. The experimental data follow the theoretical values from Eq.1 (Red lines), which is also consistent with the previous results[3]. Particularly, the resonance frequencies $v_1$ and $v_2$ mix together at the magic angle $\theta$ of 54.7 º [3,31]. The ODMR measurements show the potential of the silicon vacancy-based fiber-integrated magnetometer for high sensitivity vector magnetic field sensing.

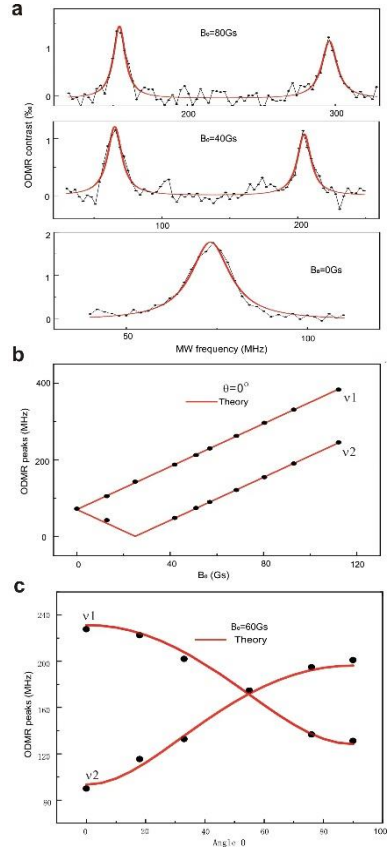

**Figure 4** Magnetic field sensing based on the fiber-integrated magnetometer. (a) Three ODMR spectra under the c-axis magnetic field of different magnitudes. The red curves are Lorentz fitting of peaks. (b) ODMR peaks as a function of magnitude of c-axis magnetic fields. The Red lines are the theoretical values. (c) ODMR peaks with respect to the angles between the silicon vacancy and magnetic field direction while keeping $B_0$ to 60 Gs. The Red lines are the theoretical values. The error bars's size is little than that of the markers.

In summary, we have demonstrated a compact fiber-integrated silicon vacancy in SiC-based magnetometer at ambient conditions. First, we realize the efficient coupling of the silicon vacancy in micron SiC with a fiber tip and control of the spin state through the fiber at the same time. By optimization of the laser and microwave power, the magnetic sensing sensitivity reaches 12.3 µT/Hz$^{1/2}$. Finally, the vector magnetometer by detecting the ODMR as a function of the strength and polar angle of a magnetic field are tested, which are in good agreement with the theory. The sensitivity can be further increased through three aspects. First, implantation of the sample with different energies can efficiently increase the silicon vacancy concentration with a larger depth. Moreover, the quenched annealing methods can simultaneously improve the counts and ODMR contrast several times[17]. Second, some methods can further enhance excitation and collection efficiency including tapered optical fiber[26], micro-concave mirror[32], and so on. Third, the pulse ODMR methods can efficiently reduce the ODMR width by decreasing the power broadening induced by laser and microwave[28,33]. Through these experiments, we prove that the compact fiber-integrated silicon vacancy magnetometer could be used in practical environments.

**Acknowledge**

This work is supported by the National Natural Science Foundation of China (Grants 61905233, 11975221, 11874361). J. F. Wang also acknowledges financial support from the Science Specialty Program of Sichuan University (Grand No. 2020SCUNL210). X. D. Liu is grateful for the support from the Youth Innovation Promotion Association of CAS (No. 2021446), and CAS Innovation Grant (No. CXJJ-19-B08)

**Notes**

The authors declare no competing financial interest.